# Intersection of low-energy electron-atom scattering and photodetachment of negative ions

Zineb Felfli and Alfred Z. Msezane
Department of Physics and Center for Theoretical Studies of Physical Systems, Clark Atlanta University, Atlanta, Georgia 30314, USA

**ABSTRACT**

We propose to use the near-threshold electron scattering data for atoms to guide the reliable experimental determination of their electron affinities (EAs), extracted using the Wigner Threshold Law, from laser photodetachment threshold spectroscopy measurements. Data from the near-threshold electron elastic scattering from W, Te, Rh, Sb and Sn atoms calculated using our complex angular momentum method, wherein is embedded the electron-electron correlations and core polarization interaction, are used as illustrations. We conclude with a remark on the relativistic effects on the EA calculation for the heavy At atom.

**Introduction**

The electron affinity (EA) of atoms, important *inter alia* in chemical reaction dynamics, including catalysis, provides a stringent test of theoretical methods when the results are compared with those from reliable measurements since the EA calculations probe the many-body effects in many electron systems. Previously, using the complex angular momentum (CAM) method [1], we calculated EAs for the lanthanide atoms [2]. Considerable disagreement was found when comparing our results to the measured EAs for the Ce, Pr, Nd, Eu, Tb, Yb and Tm atoms (see Table II of [2]). Interestingly, the recently obtained EA value for the complex Ce atom from the combined tunable infrared laser photodetachment spectroscopy and Relativistic Configuration Interaction Continuum formalism [3] is now converging to our CAM-calculated EA value [2, 4].

Generally, the near-threshold electron elastic TCSs for atoms is characterized by Ramsauer-Townsend (R-T) minima, shape resonances and, if they exist, bound (ground and excited) states of the negative ions formed during the collision of the incident electron with the target atoms as long-lived resonances [5]. The presence of these resonances could be problematic in the use of the Wigner Threshold Law, yielding results that are uncertain. This is particularly relevant for photodetachment measurements which are now highly sophisticated but ultimately rely on the Wigner Threshold Law in the determination of the bound state energies of negative ions by extrapolating their photodetachment cross sections in the near-threshold region to zero.

Moreover, since the Wigner Threshold Law [6] is only the first term of a power series expansion in the electron momentum, deviations occur when higher-order terms become important [7, 8]. Short range interactions such as induced polarizations, static multipole moments of the residual atom and the existence of nearby Feshbach resonances affect the range of validity of the Wigner Threshold Law for negative ions [9] and this has been discussed in the context of the iodine negative ion [8]. The weakness of the Wigner Threshold Law when fitting the measured data for the Ir⁻ and Pt⁻ ions has been discussed as well [10, 11] and the observed significant systematic deviation from it of the spectra of both the Ir and Pt atoms was noted. Significantly, Hotop et al [12] found that in the case of the Se⁻ ion departures from the Wigner's leading term behavior of the cross section occurred as close as 5meV above threshold.

In this paper we demonstrate the difficulties that could be encountered, when using the Wigner Threshold Law to extrapolate the measured photodetachment cross section of an atomic negative ion without considering the rich structure of the near-threshold cross section for the corresponding electron-neutral atom scattering. Note that the two processes are the reverse of each other. It is therefore envisioned that the mapping and understanding of the energy region of the intersection of the near-threshold electron scattering by atoms and laser photodetachment of negative ions could be useful in guiding the reliable determination of electron affinities of atoms from laser photodetachment measurements of negative ions using the Wigner Threshold Law, particularly where the electron elastic TCSs are characterized by complex resonance structures.

One reason why the CAM method is very effective in describing atomic collisions is that the eigenfunctions of the Schrödinger equation, known as Sturmian functions, diagonalize the potential and therefore are superior to energy eigenstates for representing waves in regions where the potential is strongest and often provide rapidly convergent expansions of wave functions for complex systems [13]. This is highlighted here by the case of the Sb atom. Using the multiconfiguration Dirac –Hartree-Fock (MCDHF) method, Li et al [14 ] calculated electron affinities for selected elements of Groups III, IV, V, VI and VII but found that for the elements Sb and Bi of group V the number of configurations was too large for the calculation of the electron affinities to be handled by their method. Using our CAM-based method, we performed the calculation for the e-Sb scattering and obtained the value of 1.258.eV for the electron affinity which compares reasonably well with the measured value of 1.047eV [15], considering the fact that our calculation is non-relativistic.

**Choice of Atoms**

To best illustrate our objective we have selected the interesting representative atoms Rh, Sn, Sb, Te and W, with Z varying from 45 through 74, and representing Groups VIII, IV A, V A, VI A and VI B, respectively. The near threshold electron scattering data for Sn, W, Rh and Te atoms were also calculated using our CAM method and the extracted EAs are compared with the measured ones. Even for the Sn, atom the calculation of the EA using structure-based methods is a formidable feat [22] since it involves calculating the binding energy differences of a 52-body system. This proposed approach promises to usher in a new theoretical framework for the generation of reliable EAs even for complex atomic systems.

We have selected the data for electron scattering from the Sn atom as one of our illustrative examples because accurate EAs, both experimental, measured mainly by Laser photodetachment electron spectrometry (LPES) and Laser-photodetachment threshold (LPT) spectroscopy, and theoretical are available [14,16 -21], including the most recently measured value by Laser photodetachment microscopy (LPM) [22]. These results and the measured EA for In using infrared photodetachment threshold spectroscopy [23] are in excellent agreement with our CAM calculated values. We note that the ground states of the Sn and In [24] negative ions appear as isolated resonances in our calculated elastic TCSs.

Recently, the electron affinity of atomic W was measured using laser photodetachment threshold spectroscopy in a collinear geometry and a bound excited state of the $W^-$ ion was identified [25]. The agreement between the EA of W [25] and (data from other measurements**)** [26, 27] was excellent. The CAM method has also been employed to investigate the near threshold electron elastic TCSs for the W atom to identify and delineate the resonances in the TCSs as well as to extract the shape resonances, R-T minima and the BEs of the negative ions (ground and excited) formed during the collisions as resonances. The resultant characteristic complex resonance structure in the TCS for the W atom and the availability of its measured EA [25-27] qualifies the W atom as the ideal system to illustrate our objective. The Rh atom is also of interest since it exhibits a rich resonance structure in the near threshold electron elastic scattering TCS and its measured EA is also available [28]. Both atomic W and Rh are well known for their importance in catalysis and application in electronics; the latter is an exceptional catalyst.

The measured EA for the Te atom is available [29, 30], including the MCDHF calculated value [14]. The resemblance to that of the Au atom of the mapped and delineated near threshold electron elastic TCS for the Te atom, particularly the appearance of the bound state of its negative ion at the second minimum and its relatively large EA (~ 2.0 eV), could qualify the Te atom and its anion as appropriate candidates for use as nanocatalysts. The reason is that the interplay between resonances and R-T minima in the electron elastic TCSs for the Au atom, along with its large EA has been advanced as the fundamental atomic physics mechanism responsible for the nanoscale catalysis [31, 32]. Finally, for the At atom with a Z of 85 we compare our extracted EA value [33] with those from the MCDHF [14] and Zollberg [34] to assess the importance/unimportance of relativistic effects. Note that all the systems presented here are relatively heavy making the calculation of their electron affinities using structure-based methods a daunting task.

**Calculation Method**

Resonances are defined rigorously as singularities of the *S*-matrix. Here, we extract unambiguous shape resonances and BEs as well as distinguish between shape resonances and the BEs of the stable ground and excited states of the negative ions thus formed during the collision as Regge resonances through the close scrutiny of the imaginary part of the complex angular momentum L, Im L. For the BE of the ground state of the resultant negative ion, Im L is several orders-of-magnitude smaller than the value corresponding to that of the attendant excited state or shape resonance [1].

The calculations here use our CAM, also known as Regge-pole, methodology [2] wherein is embedded the crucial electron-electron correlations and the vital core polarization interaction, which are responsible for the existence and stability of typical negative ions. Within the CAM representation of scattering, the Mulholland formula for the electron elastic TCS takes the form [2] (atomic units are used throughout):

$$\sigma_{tot}(E) = 4\pi k^{-2} \int_0^\infty \text{Re}[1 - S(\lambda)]\lambda d\lambda$$
$$- 8\pi^2 k^{-2} \sum_n \text{Im} \frac{\lambda_n \rho_n}{1 + \exp(-2\pi i \lambda_n)} + I(E) \quad (1)$$

where S is the S-matrix, k = √(2mE), with m being the mass, $\rho_n$ the residue of the S-matrix at the nth pole, $\lambda_n$ and I(E) contains the contributions from the integrals along the imaginary λ-axis; its contribution has been demonstrated to be negligible [1]. We consider the case for which Im $\lambda_n$<<1 so that for constructive addition, Re $\lambda_n \approx$ 1/2, 3/2, 5/2, yielding $\ell = \text{Re L} \cong 0,1,2...$ . The importance of Eq. (1) is that a resonance is likely to affect the elastic TCS when its Regge pole position is close to a real integer.

For the calculations we use the Thomas-Fermi type model potential in the well investigated form [35, 36]

$$U(r) = \frac{-Z}{r(1 + aZ^{1/3}r)(1 + bZ^{2/3}r^2)}, \quad (2)$$

where $Z$ is the nuclear charge and *a* and *b* are adjustable parameters. For small r, the potential describes the Coulomb attraction between an electron and a nucleus, $U(r) \sim -Z/(r)$, while at large distances it mimics the polarization potential, $U(r) \sim -1/(abr^4)$ and accounts properly for the vital core-polarization interaction at very low energies. The effective potential

$$V(r) = U(r) + L(L+1)/(2r^2), \quad (3)$$

is considered here as a continuous function of the variables $r$ and L. The details of the calculations are found in [2].

The R-T minima are manifestations of the polarization of the atomic core by the scattered electron [37]. Generally, the low-energy electron elastic scattering TCSs for metal atoms in their ground states are characterized by two R-T minima, one just before the shape resonance and the second one appearing immediately after the shape resonance (Figs. 2 and 4 below are typical examples). These minima are important in understanding the process of sympathetic cooling and the production of cold molecules from natural fermions [38] as well as nanocatalysis through atomic negative ions [31, 32]. Reliable atomic and molecular affinities, manifesting the existence of long-lived negative ion formation, are crucial to the understanding of a large number of chemical reactions involving negative ions [39]. The role of resonances is to promote anion formation through electron attachment [40].

**Results**

Regge trajectories allow us to probe electron attachment at its most fundamental level near threshold, thereby uncovering new manifestations such as the origin of resonances (shape resonances, ground and excited states of negative ions) formed during the slow electron collisions. They also lead to a fundamental understanding of nanocatalysis through negative ion resonances and the extraction of the reliable electron affinities. In Fig. 1 we present detailed results for the electron scattering from the Sn atom, demonstrating how the various resonances originate from the examination of the Regge trajectories. For the subsequent atoms W, Te, Sb and Rh we show only the data corresponding to Fig. 1 (a); the rest is omitted to keep the paper both focused and compact.

Figure 1(a) presents the elastic TCS (a.u) and the Mulholland partial cross sections (a.u) versus $E$ (eV) for $e$ - Sn scattering. The TCS is characterized by a maximum near threshold followed by a R-T minimum at about 0.087 eV and two additional resonances at 0.312 and 1.10 eV, with Re $L$=2 and Re $L$=4, respectively. The corresponding Im $L$ values are, respectively, 4.40 x$10^{-2}$ and 1.3 x$10^{-4}$. Clearly, the latter corresponds to a long-lived state, a bound state of the Sn⁻ negative ion since the angular life is proportional to 1/ (Im $L$) [41], while the former represents a short-lived resonance; in this case it is a shape resonance. The R-T minimum at 0.087 eV is generated through the interference between the $n$=3 and $n$ =4 Mulholland partial cross sections; the latter also determines the Wigner threshold behavior.

Figure 1(b) displays the magnified shape resonance at 0.312 eV, corresponding to Re $L$=2 and Im $L$=0.044. Note the significant difference between the angular lives of the shape resonance and the stable bound state of the Sn⁻ negative ion (see also Fig. 1(d)). Figure 1(c) displays the Regge trajectories, demonstrating that indeed the $n$=3 trajectory at Re $L$=2 is responsible for the resonance at 0.312 eV, while the $n$=0 trajectory with Re $L$=4 accounts for the very sharp resonance at 1.10 eV; these trajectories are close to integer values of Re $L$. We note that this is a long-lived resonance as seen from its large angular life 1/(Im $L$) [41] and corresponds to the EA for the Sn atom. A closer look at the bound state of the Sn⁻ negative ion is given in the expanded Fig. 1(d). The energy value for this resonance should be compared with the measured and calculated EAs [14, 16-22].

Figure 2 compares and contrasts the electron elastic TCSs for the W ground state atom (red curve) and its excited state (blue curve). The ground state TCS curve is typical; it is characterized by two R-T minima and a shape resonance which is followed by the dramatically sharp resonance at the second R-T minimum of the cross section. The position of this resonance corresponds to the BE of the W⁻ ion formed in the ground state during the collision; it is identified as the EA of the W atom. Note that the SR at 0.86 eV and the EA at 1.59 eV appear quite close together. The blue curve, typical of the electron elastic TCS for an excited state (see for example Ref. [5] figure 1(d) for the Au TCSs), exhibits the sharp resonance at 0.61 eV which corresponds to the BE of an excited state of the W⁻ ion. And it sits at the first R-T minimum of the TCS for the ground state and is preceded by a SR at about 0.28 eV. Most interesting about the comparison is the appearance of the BE of the excited state of the W⁻ ion and the EA of the W atom at the first and the second R-T minima of the TCS, respectively.

Figure 3 compares the electron elastic TCSs for the Rh ground state (red curve) and the excited state (blue curve). Note the appearance of the bound state of the Rh⁻ ion at the second R-T minimum of the ground state TCS curve of Rh and that the first and the second R-T minima have almost the same depth. This first such observation for the Rh atom and the configuration of the R-T minima as well as the large EA of Rh are probably connected with the atomic Rh's excellent catalytic properties.

In Figure 4 is contrasted the electron elastic scattering cross sections for the ground and the excited states of the Te atom in the energy range E< 7 eV; these results are similar to those for the W atom. Our calculated EAs of the Rh, Sn, Sb, Te and W atoms are summarized in the Table 1, where they are compared with other measured and calculated values. Included are also the theoretical EAs for the At atom and the corresponding shape resonances, Ramsauer-Townsend minima and binding energies for the ground and the excited Rh⁻, Sn⁻, Sb⁻, Te⁻ and W⁻ ions, all in eV. As seen from Table 1, for the atoms Sn and Te the agreement between the measured EAs and our calculated data is within 1% and 11%, respectively. However, for the W atom our SR and the measured EA [25] are very close together, namely 0.86 eV and 0.816 eV, respectively. Furthermore, for the excited Rh⁻ ion our calculated BE is 0.35 eV while the measured value is less than 0.385 eV [10]. Our SR for Rh is 1.70 eV while the measured EA is 1.143 eV [10] (this should be compared with our calculated EA of 3.12 eV. Consequently, these discrepancies call for immediate investigations, both experimental and theoretical for resolution. Similar discrepancies have been identified recently for the EA of the Os atom [42]. It is noted that the near-threshold electron elastic TCSs for the Os and W atoms are similar; they are characterized by a rich resonance structure.

**Summary and Conclusions**
In this paper we have explored using the CAM methodology the energy region of the intersection of the low-energy electron elastic scattering from atoms with the photodetachment of negative ions. In this region slow electrons attach to atoms forming stable negative ions as resonances; negative ions are laser photodetached and the Wigner Threshold Law is employed to obtain the EAs. Electron elastic scattering data, including new TCSs, for the Sn, W, Rh, Sb and Te atoms have been calculated and presented as illustrations. We find that in this energy region, the electron elastic TCSs are characterized by the important R-T minima, SRs and BEs of the ground and the excited negative ions formed during the slow electron collisions as resonances; these are identified and extracted as well.

Through the Reggie trajectories we probed the electron attachment mechanism, leading to the formation of negative ions (both ground and excited) as resonances. From the resonances we extracted the unambiguous EAs of the relevant atoms through the close scrutiny of Im L, obtaining very good agreement with the measured values for the Sn and Te atoms. It is concluded that these resonances could be problematic when using the Wigner Threshold Law to extract the EAs from the experimental data obtained from laser photodetachment of negative ions. Interestingly, the EA of the Rh atom is calculated to be 3.12 eV, contrary to the measured value of 1.143 [10]. This large calculated EA value is consistent with the excellent catalytic property of the Rh atom.

We conclude by recommending the exploration both experimentally and theoretically of the resonance rich near threshold energy region, particularly for the case of the W and Rh atoms to confirm the present findings. In this region the electron attachment to atoms occurs forming stable negative ions, both ground and excited. This is the energy region where the low-energy electron elastic scattering intersects the photodetachment of negative ions accessed through the use of the Wigner Threshold Law. We further conclude from the comparison of the calculated EA of the At atom using the MCDHF method [14] and our nonrelativistic CAM method that the contribution from relativity is less than about 4%. This insignificance of relativistic effects was found also in the investigation of nanoscale catalysis using negative ions [43].

**Acknowledgments**
Research was supported by U.S. DOE Office of Basic Energy Sciences, Atomic Molecular and Optical Sciences Program (Grant DE-FG02-97ER14743) and Army Research Office (Grant W911NF-11-1-0194). The calculations were performed at the NERSC computational facility.

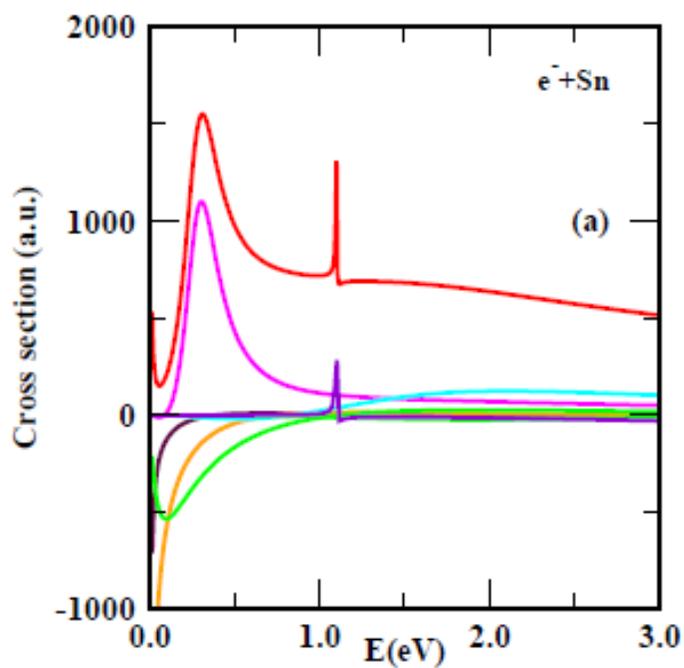
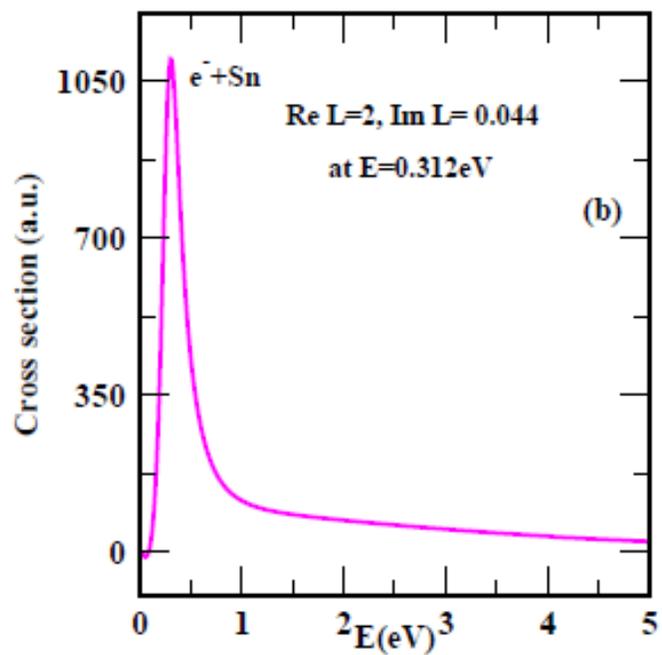
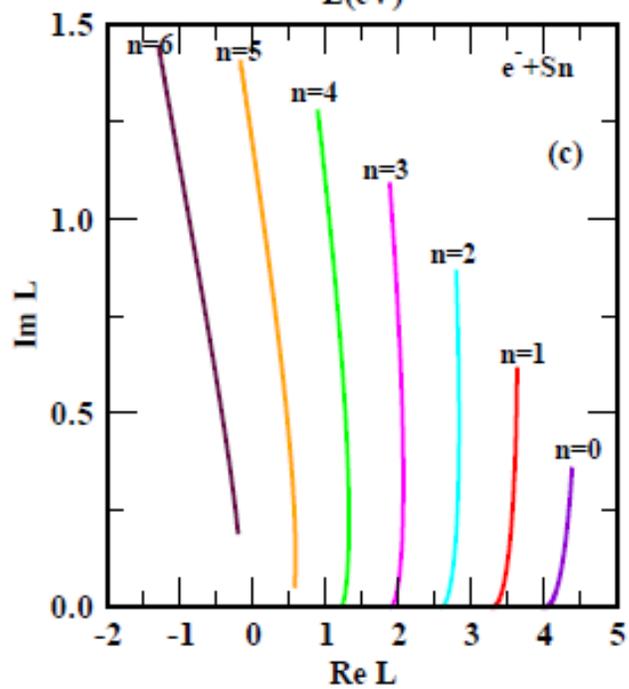
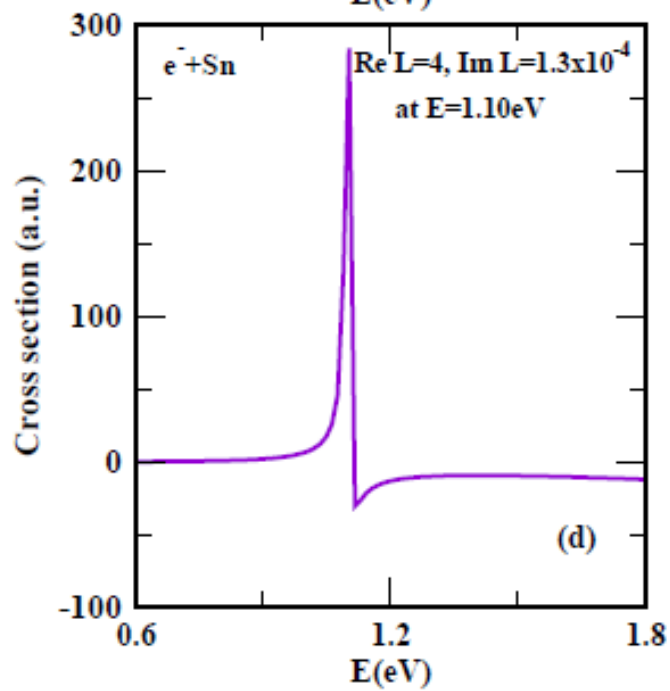

**Figure 1**

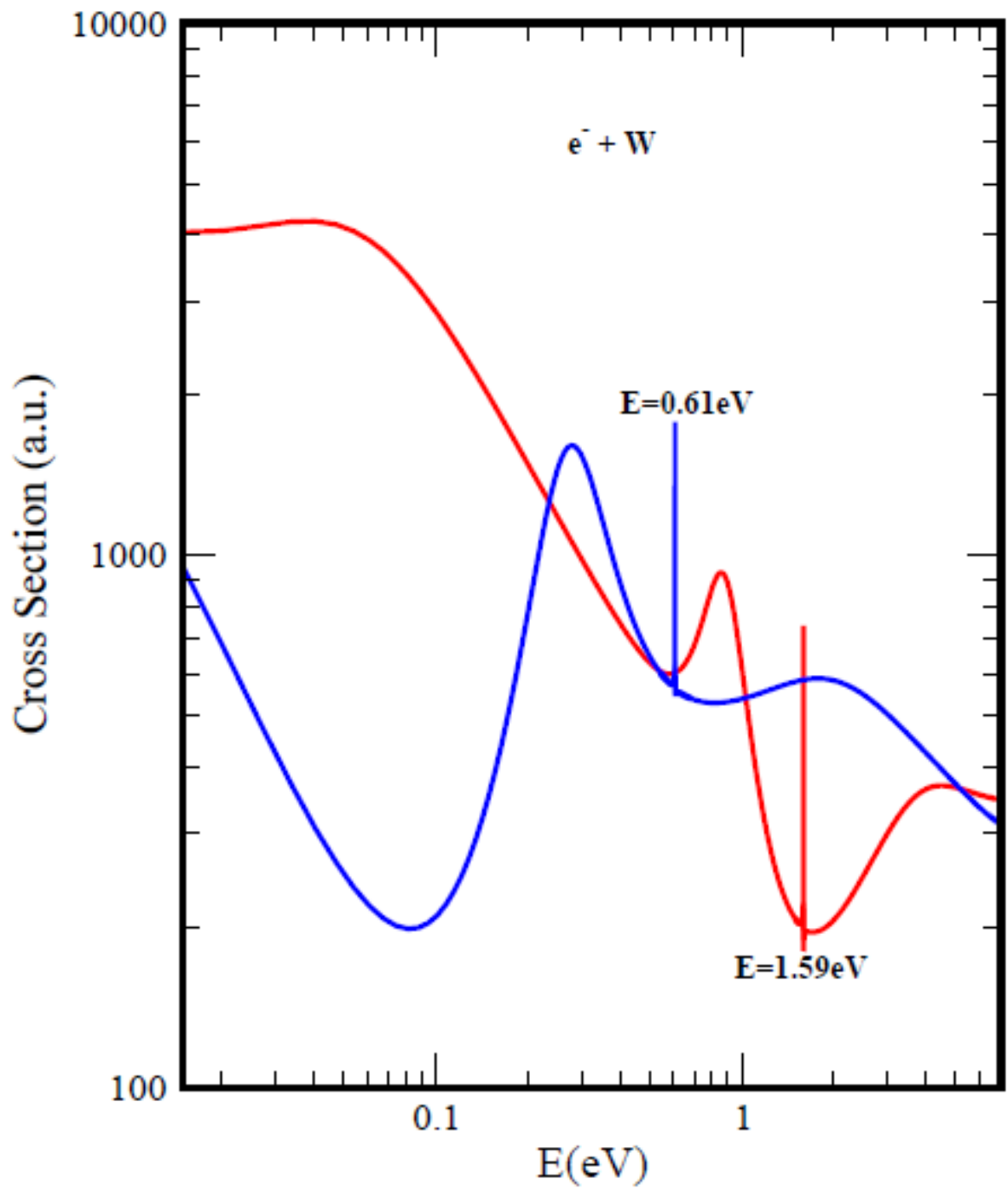

**Figure 2**

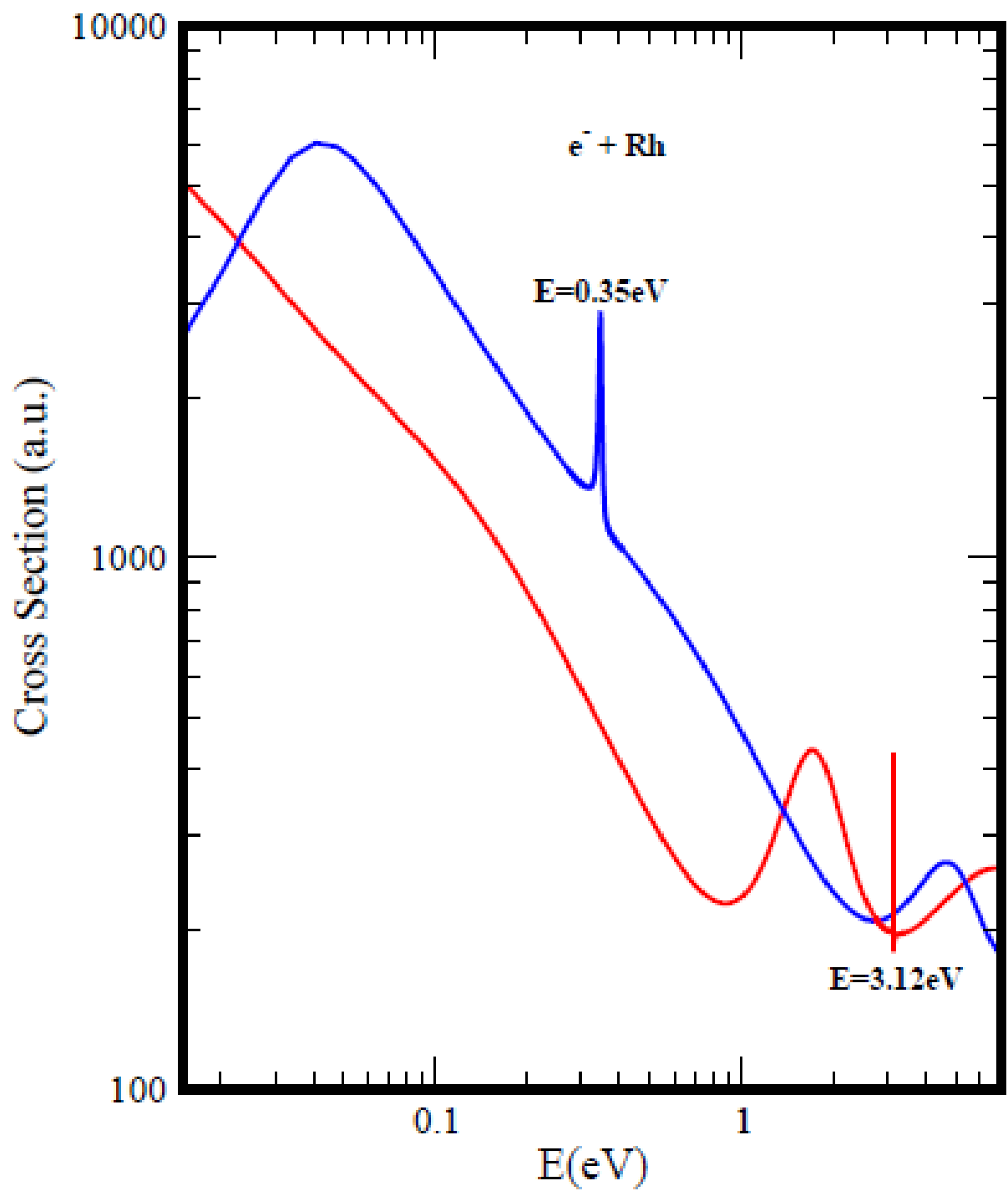

**Figure 3**

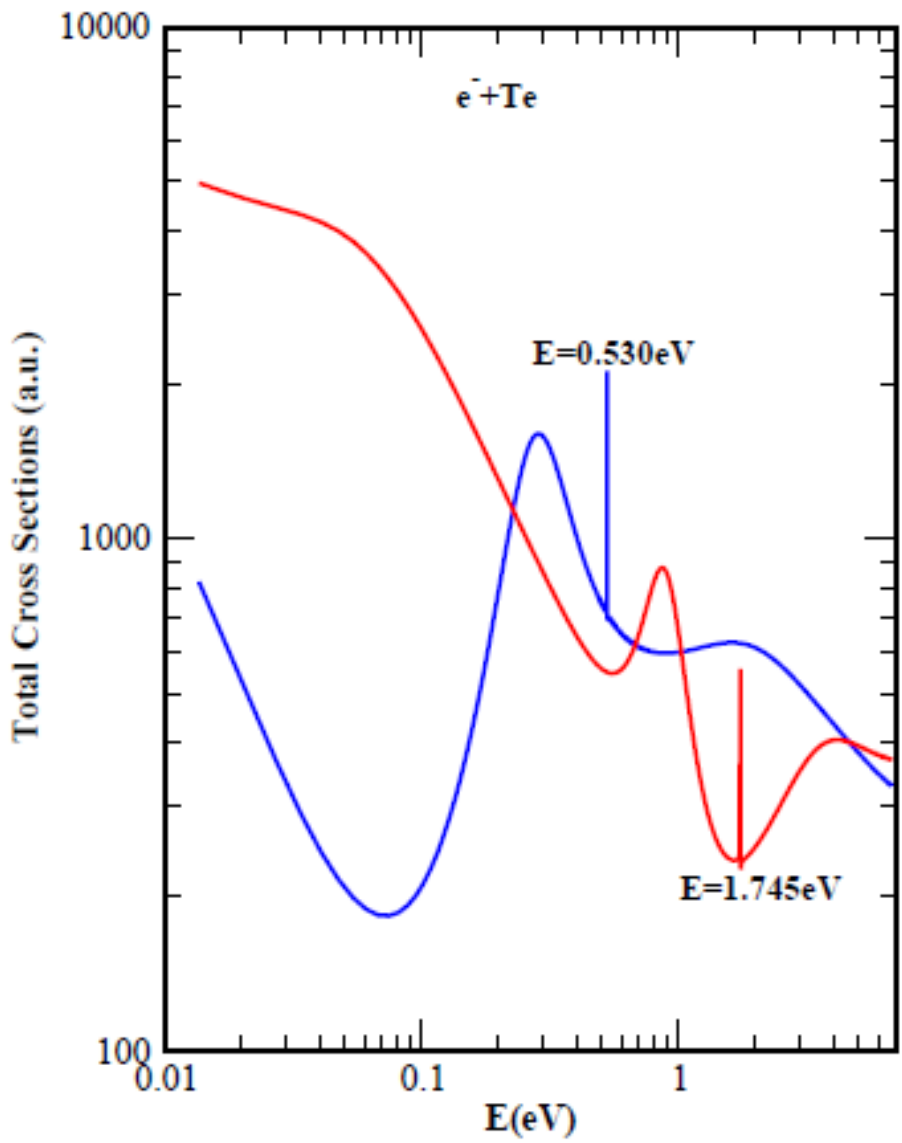

**Figure 4**

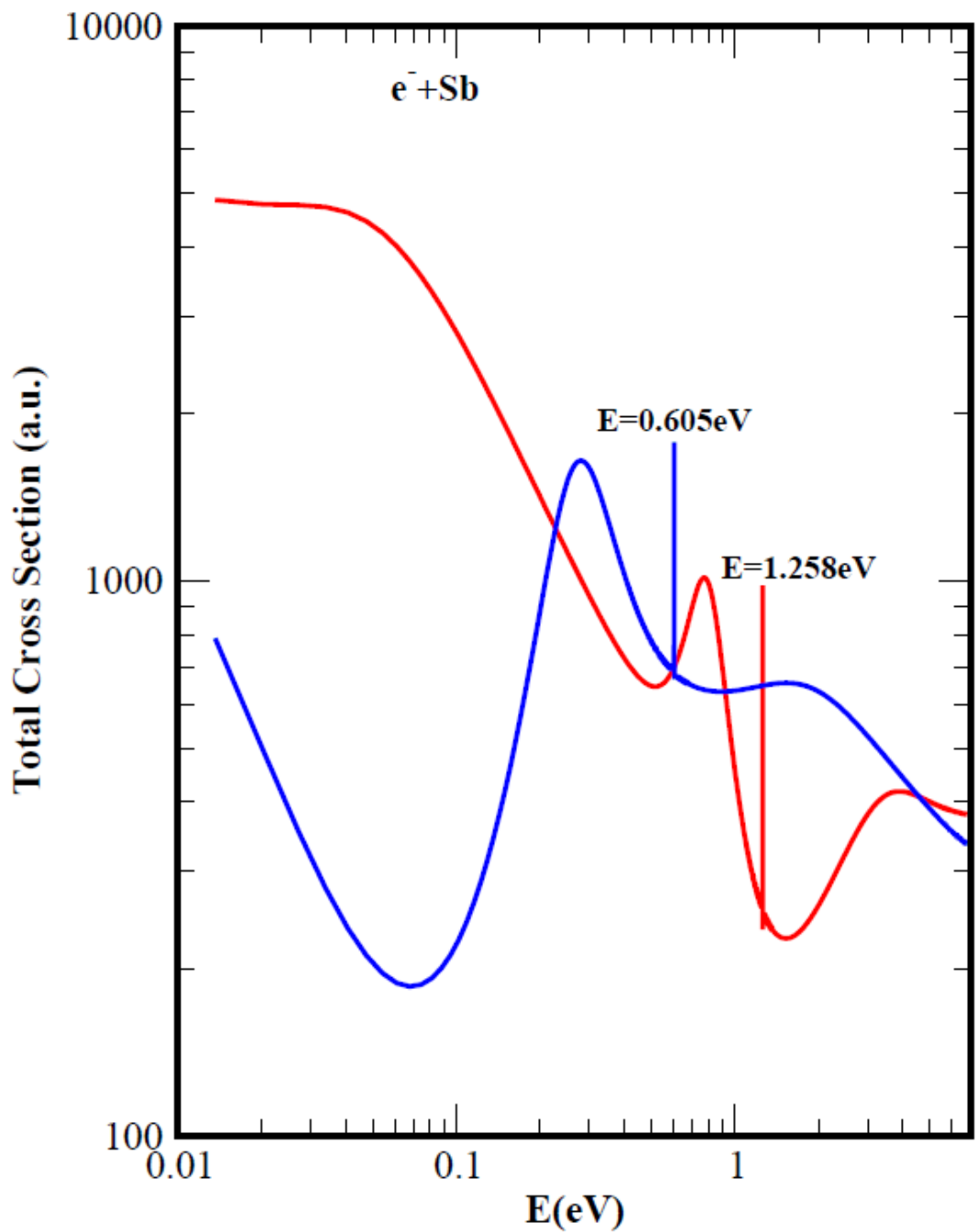

**Figure 5**

## Figure Captions

**Fig. 1:** (a) Total and Mulholland partial elastic cross sections, a.u, for *e*-Sn scattering versus $E$ (eV), showing the Mulholland contributions. The *n*=4 (green curve) Mulholland partial cross section also determines the Wigner threshold behavior. The R-T minimum is clearly visible near threshold. (b) Mulholland contribution to the TCS, a.u, for *e*-Sn scattering versus $E$ (eV), corresponding to the Regge trajectory that passes near Re $L$=2 at $E$=0.312 eV, and hence responsible for the short-lived resonance (Im $L$=0.044) in the TCS at that energy. (c) Regge trajectories, viz., Im $L$ ($E$) versus Re $L$ ( $E$), for *e*-Sn scattering, demonstrating the main Re $L$ ($E$) contributors to the TCS. (d) Mulholland contribution to the TCS, a.u, for *e*-Sn scattering versus $E$ (eV), corresponding to the Regge trajectory that passes near Re $L$=4 at $E$=1.10 eV, and hence responsible for the resonance (Im $L$=1.3 x10$^{-4}$) in the TCS at that energy. This is a long-lived resonance as seen from its large angular life (Im $L)^{-1}$ and corresponds to the EA for the Sn atom.

**Fig. 2**: Comparison between the electron elastic TCSs for the W ground state (red curve) and the excited state (blue curve). Note the rich resonance structure and the appearance of the bound states of the W¯ ion at the first (excited ion) and second (ground state ion) R-T minima of the ground state TCS of the W atom. Also note the significant difference between the Figs. 1(a) and 2 for the ground state TCS curves.

**Fig. 3**: Comparison between the electron elastic TCSs for the Rh ground state (red curve) and the excited state (blue curve). Note the appearance of the large bound state (3.12 eV) of the Rh¯ ion at the second R-T minimum of the ground state TCS curve of the Rh atom and that the first and the second R-T minima have almost the same depth.

**Fig. 4**: Comparison between the electron elastic TCSs for the Te ground state (red curve) and the excited state (blue curve). Note the appearance of the bound states of the Te¯ ion at the first (excited ion) and second (ground state ion) R-T minima of the ground state TCS curve of the Te atom.

**Fig. 4**: Comparison between the electron elastic TCSs for the Sb ground state (red curve) and the excited state (blue curve). Note the appearance of the bound states of the Sb¯ ion at the first (excited ion) and second (ground state ion) R-T minima of the ground state TCS curve of the Sb atom.

**Table 1:** Electron affinities (EAs), Shape Resonances (SRs), Ramsauer-Townsend (R-T) minima and Binding energies (BEs) of the excited negative ions, all in eV, for Rh, Sn, Sb, Te, W and At atoms. Extd, Expt. and Present denote excited state, experiment and present calculation, respectively.

| Atom | Z | 1st R-T Minimum | SR | 2nd Minimum | EA Present | EA Expt. | BE (Extd) Present | BE(Extd) Expt. |
|---|---|---|---|---|---|---|---|---|
| Rh | 45 | 0.88 | 1.70 | 3.06 | 3.12 | 1.143 [10] | 0.35 | < 0.385 [10] |
| Sn | 50 | 0.054 | 0.312 | -- | 1.100 | 1.112 [22] | 0.285 | |
| Sb | 51 | 0.516 | 0.774 | 1.250 | 1.258 | 1.047 [15] | 0.605 | |
| Te | 52 | 0.557 | 0.869 | 1.656 | 1.745 | 1.973 [29] | 0.530 | |
| W | 74 | 0.58 | 0.86 | 1.56 | 1.59 | 0.816 [25] | 0.61 | |
| At | 85 | 0.810 | 1.200 | 2.490 | 2.51 | 2.80 [34 ]* 2.415[14 ]* | 0.292 | |

* Theoretical values. The MCDHF [14] calculated EA value deviates from the CAM calculated one by about 3.8 %.